\documentclass[pra,twocolumn,tightenlines,showpacs,nofootinbib]{revtex4}
\usepackage{bm,dcolumn,amsmath,graphicx,amsfonts,amssymb}
\usepackage{epsfig}

\begin{document}
\title{Quantum electrodynamics corrections to energies, transition
  amplitudes and parity nonconservation in Rb, Cs, Ba$^+$, Tl, Fr and Ra$^+$.} 
 
\author{B. M. Roberts$^1$, V. A. Dzuba$^1$, and V. V. Flambaum$^{1,2}$}

\affiliation{$^1$School of Physics, University of New South Wales,
Sydney, NSW 2052, Australia}

\affiliation{$^2$Centre for Theoretical Chemistry and Physics, New Zealand
Institute for Advanced Study, Massey University, Auckland 0745, New Zealand}

\date{ \today }

\begin{abstract}
We use previously developed radiative potential method to calculate
quantum electrodynamic (QED) corrections to energy levels and electric
dipole transition amplitudes for atoms which are used for the study of
the parity non-conservation (PNC) in atoms. The QED shift in energies
and dipole amplitudes leads to noticeable change in the PNC
amplitudes. This study compliments the previously considered QED
corrections to the weak matrix elements. We demonstrate that the QED
corrections due to the change in energies and dipole matrix elements
are comparable in value to those due to change in weak matrix
elements. 
\end{abstract}
\maketitle

\section{Introduction}

The study of parity nonconservation (PNC) in atoms provides a
low-energy search for new physics beyond the standard model (see,
e.g.~\cite{GFrev04,DFrev12}), which is a relatively inexpensive
alternative to the high-energy searches performed in colliders.  
In fact, PNC in cesium is currently the most precise low-energy test
of the electroweak theory due to the high accuracy of the
measurements~\cite{meas1,meas2} and the calculations needed for their
interpretation~\cite{CsPNC} (see
also~\cite{RbPNC,CsPNCour1,CsPNCour2,CsPNCold}). 

The level of precision that has been obtained in atomic physics
calculations and measurements has meant that strong-field QED
corrections are observable and must be taken into account. 
It was shown in fact, that the inclusion of self-energy-type QED
corrections to PNC calculations in cesium restored the agreement with
the standard model~\cite{KFQED03,MSTQED03,ueh} (see
also~\cite{otherQED,FGradpot,Shabaev}). Just as these calculations
have proven to be important in the case of cesium, they will be
necessary for the calculations of other atoms as the accuracy
increases and as new experiments become available.   
For this reason, using the ``radiative potential'' method developed in
Ref.~\cite{FGradpot}, we present calculations of the QED corrections to the
PNC amplitudes of several transitions in Rb, Cs, Ba$^+$, Tl, Fr and
Ra$^+$. 

The case of rubidium is interesting because of its simple electron
structure and small value of different corrections to the PNC
amplitude~\cite{RbPNC}. The interpretation of the PNC measurements for
Rb can be more reliable than for other atoms. 
We have shown in our previous work~\cite{RbPNC} that the accuracy of
the calculations for rubidium can surpass those for cesium, while the
PNC amplitude is only several times smaller~\cite{RbPNC} (see
also~\cite{DzubaTl}). 

Francium is a particularly important application. The FrPNC
collaboration has begun the construction of a laser cooling and
trapping apparatus at  the TRIUMF laboratories in Canada with the
purpose of measuring atomic parity nonconservation  in artificially
produced francium~\cite{FrPNC}. With a PNC amplitude expected to be
around 15 times larger than that of cesium, and its relatively simple
electronic configuration which leads to accurate calculations,
francium is an  ideal atom for precision measurements of
PNC~\cite{FrDzuba,Ba+,FrSaf}. 
When these measurements become available it will be very important to
have accurate atomic calculations for analysis, and these calculations
will require contributions from quantum electrodynamics effects.   

There are accurate calculations and measurements available for
thallium~\cite{DzubaTl,KozlovTl,Fortson}, and measurements have also
been considered for the Ba$^+$ ion~\cite{Ba+,BaII} and are in progress
for the Ra$^+$ ion~\cite{KVI}. 

%==============================================================
\section{QED corrections} % -- %\section{Methods} 

The quantum electrodynamics corrections considered
in this work arise from vacuum polarization and elec-
tron self-energy. The inclusion of vacuum polarization
is numerically relatively simple, achieved via inclusion of
the Uehling potential. The self-energy contribution is in-
cluded via the radiative potential method developed in
Ref.~\cite{FGradpot}. Note that the radiative correction to the electric dipole
transition operator $d_{E1}$ (vertex correction) is very small and may
be neglected. The  change of the electric dipole matrix elements come
from the QED corrections to the electron wave functions. 

The parity nonconservation amplitude of a transition  ($a-b$) between
states of the same parity can be expressed via the sum over all
possible intermediate opposite parity states $n$, 
%\begin{align}
\small
\begin{equation}
E_{PNC} = \sum_n   %&
	 \Big[ \frac{\langle b|\hat{d}_{E1}|{n}\rangle\langle{n}|\hat{h}_{W}|{a}\rangle}{E_{a}-E_{n}}  % \notag\\&\qquad
	+\frac{\langle b|\hat{h}_{W}|n\rangle\langle n|\hat{d}_{E1}|{a}\rangle}{E_{b}-E_{n}}\Big],
	\label{eq:pnc}
\end{equation}
\normalsize
%\end{align}
where $\hat{d}_{E1}$ is the electric dipole transition operator and
$\hat{h}_{W}$ is the nuclear-spin-independent weak interaction.  

We consider QED corrections from the three sources separately --
corrections to the weak amplitudes, the dipole amplitudes and the
energy denominators.   

In the original works, QED corrections were only calculated  for the
weak matrix elements (see e.g.~\cite{KFQED03,MSTQED03,ueh,otherQED}).
This was a reasonable approximation numerically for cesium due to a
chance cancellation between QED contributions coming from corrections
to the energy levels and dipole matrix elements, which were calculated
by Flambaum and Ginges~\cite{FGradpot}. This cancellation  is not
guaranteed    and it was demonstrated in Ref.~\cite{FGradpot} that
corrections from all three sources are equally important and must be
included. 
Full determinations of QED corrections to the entire PNC amplitude
have only been considered for cesium
\cite{CsPNCour2,FGradpot,Shabaev}.   

The radiative potential method, developed by Flambaum and
Ginges~\cite{FGradpot}, defines an approximate potential $\hat{L}$
such that the radiative correction to the energies coincides with its
average value, 
\[\delta E_n = \langle n|\hat{L}|n\rangle.\]
This potential takes into account the local vacuum polarization
operator (including the lowest-order in $Z\alpha$ Uehling potential as
well as the higher order Wichmann-Kroll potential) and the non-local
strong Coulomb field electron self-energy operator. 

In this work, we use the existing calculations of QED corrections to
the weak matrix elements, which are presented in
Table~\ref{tab:weak}. 
These values have been taken from the works
Ref's~\cite{KFQED03,MSTQED03,ueh} (see
also~\cite{otherQED}). Midpoints and uncertainties have been chosen to
agree with different previous determinations. Note also that these
calculations are valid only for $s$-$p_{1/2}$ weak transitions. 

We then use the radiative potential method outlined above, with the
exception of the small Wichmann-Kroll term, to calculate the QED
corrections to the energy levels and electric dipole matrix elements.
We then calculate the dominating terms in equation (\ref{eq:pnc}) with
and without the radiative corrections to determine the total
percentage correction to the PNC amplitudes for several transitions in
Rb, Cs, Ba$^+$, Tl, Fr and Ra$^+$. 

  \begin{table}
    \centering
    \caption{Values (from Ref's \cite{KFQED03,MSTQED03,ueh}) for the
      percentage contributions of vacuum polarization including the
      Uehling and smaller Wichmann-Kroll (W-K) potentials, and
      self-energy-vertex (SE-V) to weak $s$-$p$ matrix elements for various
      atoms. Uncertainty is estimated from the spread of values of
      different sources.} 
\begin{ruledtabular}
      \begin{tabular}{lrrrr}
            { } & {Uehling} & {SE-V} & {W-K} & {Total}  \\
      \hline
      {Rb} 		& 0.20   & -0.51  & 0.001 & -0.31(2) \\
      {Cs} 		& 0.40   & -0.84  & 0.003 & -0.43(3) \\
      {Ba${}^+$} & 0.41    & -0.86  & 0.003 & -0.45(3) \\
      {Tl} 		& 0.93   & -1.44  & 0.015 & -0.50(5) \\
      {Fr} 		& 1.13   & -1.75  & 0.02 & -0.60(7) \\
      {Ra${}^+$} & 1.17   & -1.80  & 0.02 & -0.61(7) \\ 
            \end{tabular}%
\end{ruledtabular}
    \label{tab:weak}%
  \end{table}%

%=============================================
\section{Calculations}

We use the sum-over-states approach to calculate the PNC amplitudes.
Relativistic Hartree-Fock (RHF) method is used to construct single-electron
orbitals and random phase approximation (RPA) is used to include the
effect of core polarization by external fields. 

Core-valence correlations are included by means of the correlation
potential method (CPM)~\cite{CPM}. The second-order correlation
potential $\hat \Sigma^{(2)}$ is calculated using many-body
perturbation theory and then used to construct the so-called Brueckner
orbitals (BO) for the external electron.
BO are found by solving the Hartree-Fock--like equations with an extra
operator $\hat \Sigma$:  
\begin{equation}
 (\hat H_0 +\hat \Sigma^{(2)} - E_n)\psi_n^{(\rm BO)}=0,
\label{eq:BO}
\end{equation}
where $\hat H_0$ is the relativistic Hartree-Fock Hamiltonian and
index $n$ denotes valence states. The BO $\psi_n^{(\rm BO)}$ and
energy $E_n$ include correlations.   

We then use a simple scaling procedure to estimate the contribution of
higher-order correlations by including a factor in front of the
correlation potential, $\lambda \hat \Sigma^{(2)}$, which is chosen to
reproduce energy levels of the lowest lying valence states.  
A separate $\lambda$ is used for each of the initial/final states, and
another is used for each set of intermediate states (e.g. $np_{1/2}$,
$np_{3/2}$).  
For the second-order correlation potential, values for the fitting
parameter typically take values $\lambda \sim0.8-0.9$. 
This fitting generally increases the accuracy of the wave-functions
and therefore the matrix elements.

The PNC amplitude is given by the expression similar to
(\ref{eq:pnc}), in which states $a$, $b$, $n$ are single-electron RHF
states, and operators $\hat{d}_{E1}$ and $\hat{h}_{W}$ are modified to
include the effect of core polarization: $\hat{d}_{E1} \rightarrow
\hat{d}_{E1} + \delta V_{E1}$, $\hat{h}_{W} \rightarrow \hat{h}_{W} +
\delta V_{W}$. Here $\delta V_{f}$ is the correction to the
self-consistent core potential due to the effect of external field $f$
($f$ is either electric field of laser light $\hat{d}_{E1}$ or weak
electron-nucleus interaction $\hat h_W$). The corrections to the core
potential are found by solving self-consistently the RPA equations for
the core states 
\begin{equation}
(\hat H_0 -\varepsilon_c)\delta \psi_c = -(\hat f +\delta V_f)\psi_c.
\label{eq:RPA}
\end{equation} 
Here $\hat H_0$ is the RHF Hamiltonian, index $c$ numerates core
states, $f$ is the operator of external field (weak or electric
dipole). 

QED corrections are included by adding the radiative potential to the
RHF Hamiltonian $\hat H_0$. This is done on the stages of calculating
single-electron RHF and Brueckner orbitals and solving the RPA equations for
the electric dipole field. We remind the reader that we don't include
QED corrections for the weak interaction. The more reliable results
for weak matrix elements are found in different approaches considered
before in Refs.~\cite{KFQED03,MSTQED03,ueh,otherQED}.

B-spline technique~\cite{B-spline} is used to construct sets of
single-electron orbitals used in the calculate $\hat \Sigma^{(2)}$ and
for summation in (\ref{eq:pnc}). The basis states used to calculate
$\hat \Sigma^{(2)}$ are the linear combinations of B-splines which are
the eigenstates of the RHF Hamiltonian $\hat H_0$ (RHF orbitals). The
basis states used in the summation (\ref{eq:pnc}) are the eigenstates
of the $\hat H_0 + \hat \Sigma^{(2)}$ Hamiltonian (Brueckner orbitals).

\begin{figure}
\centering
\epsfig{figure=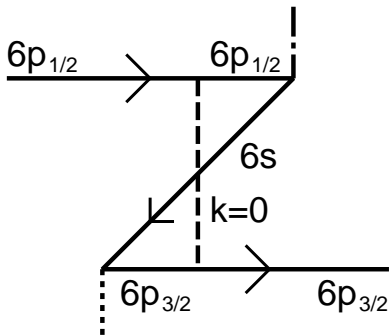,scale=1.0}
\caption{Sample contribution to the $6p_{1/2} - 6p_{3/2}$ PNC
  transition in thallium arising from the double core polarization by weak
  (dashed-dot line) and electric dipole (dots line) fields. The $6s$
  state is treated as a core state.}
\label{f:tl}
\end{figure}

The approach described above does not take into account the
effect of core polarization due to simultaneous action of the weak and
electromagnetic interactions. An example of the contribution of this
type is presented on Fig.~\ref{f:tl} for the case of thallium. The
contribution of the double core polarization to all PNC amplitudes 
considered in present paper except thallium does not exceed one per
cent. Therefore, it can be neglected in calculating the relative
effect of QED corrections. The contribution of the double core
polarization to the $6p_{1/2} - 6p{3/2}$ PNC amplitude of thallium as
about 30\% (Fig.~\ref{f:tl}) if thallium is treated as a
single-valence-electron atom, leaving the $6s$ electrons in the core. 
Note that the double core polarization was taken into account in our
old calculations of the PNC in thallium~\cite{DzubaTl} by means of the
RPA-like approach with two operators of external field. Similar
calculations in present work would not be practical since we want to
separate the effect of QED corrections on weak and electromagnetic
matrix elements and it would be hard to do so in the RPA-like approach
with two operators. Therefore, we use the configuration interaction
(CI) method instead. The calculations for Tl are similar to our recent
calculations of the EDM enhancement factor for
thallium~\cite{TlEDM}. The Tl atom is treated 
as a three-valence-electrons systems. The calculations are done in the
$V^{N-3}$ approximation~\cite{TlEDM}. The correlations between $6s$
and $6p$ electrons are included very accurately in the framework of
the CI method. The core-valence correlations are also included by
means of the many-body perturbation theory. The QED corrections are
included in electric dipole matrix elements while weak matrix elements
are kept unaffected. 

Our calculations, which include core-polarization and (fitted)
second-order Brueckner-type electron correlations, agree with existing
calculations and experimental values to around 1\% for energy levels
and a few percent for $E1$ matrix elements.  Importantly, we calculate
the PNC amplitudes which also agree well with previous calculations. A
summary of this comparison is presented in Table \ref{tab:PNC}.

  \begin{table}
    \centering
    \caption{Calculations of PNC amplitudes for several transitions in
      the RPA and fitted second-order Brueckner (BO) approximations
      and comparison with previous calculations. Values given in units
      $i (-Q_W/N)\times10^{-11}$ a.u.\,.} 
\begin{ruledtabular}
      \begin{tabular}{llll}
              \multicolumn{2}{c}{Transition}& {This work} & {Previous}  \\
             && {(RPA + BO)} & {calculations}  \\
      \hline
&&&\\
      ${}^{85}${Rb}& $5s$-$6s$ & 0.139 & 0.139 \cite{RbPNC} \\ %Use our paper!
      & $5s$-$4d_{3/2}$ & 0.449 & ---  \\ %Exist? %Should I present my calculations here? (If so, add a line to intro/abstract)
&&&\\
      ${}^{133}${Cs} & $6s$-$7s$ & 0.897 & 0.8977 \cite{CsPNC}  \\ %Use our paper!
       & $6s$-$5d_{3/2}$ & 3.75 & 3.76 \cite{Ba+}  \\
&&&\\
      ${}^{138}${Ba${}^+$}&  $6s$-$7s$ & 0.671 & --- \\ %Exist?
      &  $6s$-$5d_{3/2}$ & 2.36  & 2.34 \cite{Ba+}  \\ 
&&&\\
%      ${}^{205}${Tl}& $6p_{1/2}$-$6p_{3/2}$ &  &   \\
%&&&\\
      ${}^{223}${Fr}& $7s$-$8s$ & 15.4   &  15.41 \cite{FrSaf}  \\ %(17)
      & $7s$-$6d_{3/2}$ & 59.4 &  59.5 \cite{Ba+}\\
&&&\\
      ${}^{226}${Ra${}^+$}& $7s$-$8s$ & 10.9 &  ---  \\ %Exist?
      & $7s$-$6d_{3/2}$ & 46.6 & 45.9 \cite{Ba+}\\
            \end{tabular}%
\end{ruledtabular}
    \label{tab:PNC}%
  \end{table}%

%==============================================================
\section{Results and discussion}

Percentage QED corrections to individual lowest energy levels are presented
in Table \ref{tab:energy}, and corrections $R_{E1}$ to dipole matrix
elements are presented in Table \ref{tab:dipole}. The factor $R_{E1}$
is defined 
\begin{equation}
\langle{b}|\hat d_{E1}|{a}\rangle =\langle{b}|\hat
d_{E1}|{a}\rangle_0(1+\dfrac{\alpha}{\pi}R_{E1}), 
\label{eq:rsp}
\end{equation}
where the subscript 0 indicates the zeroth order matrix element,
without radiative corrections, and $\alpha$ is the fine structure
constant. 

  \begin{table}
    \centering
    \caption{Percentage QED corrections to ionization energies of
      lowest states for several atoms.}
\begin{ruledtabular}
\begin{tabular}{l|rrrrrrr}
            Atom&\multicolumn{7}{c}{Correction (\%)} \\
      \hline
Rb    & $5s$  & $6s$  & $5p_{1/2}$ & $6p_{1/2}$ & $7p_{1/2}$ & $4d_{3/2}$ & $5p_{3/2}$ \\
      & -0.040 & -0.023 & 0.003 & 0.002 & 0.002 & 0.003 & 0.001 \\
      &       &       &       &       &       &       &  \\
Cs    & $6s$  & $7s$  & $6p_{1/2}$ & $7p_{1/2}$ & $8p_{1/2}$ & $9p_{1/2}$ & $5d_{3/2}$ \\
      & -0.069 & -0.040 & 0.006 & 0.004 & 0.003 & 0.002 & 0.031 \\
      &       &       &       &       &       &       &  \\
Ba${}^+$ & $6s$  & $7s$  & $6p_{1/2}$ & $7p_{1/2}$ & $8p_{1/2}$ & $9p_{1/2}$ & $5d_{3/2}$ \\
      & -0.055 & -0.035 & 0.005 & 0.004 & 0.003 & 0.002 & 0.028 \\
      &       &       &       &       &       &       &  \\
Tl\footnotemark[1]    & $6p_{1/2}$ & $6p_{3/2}$ & $6s$  & $7s$  &   &  &  \\
      & -0.01 & -0.02 & -0.14 & -0.07 &  &  &  \\
      &       &       &       &       &       &       &  \\
Fr    & $7s$  & $8s$  & $7p_{1/2}$ & $8p_{1/2}$ & $6d_{3/2}$ & $7p_{3/2}$ & $8p_{3/2}$ \\
      & -0.142 & -0.076 & 0.002 & 0.001 & 0.067 & -0.003 & -0.002 \\
      &       &       &       &       &       &       &  \\
Ra${}^+$ & $7s$  & $8s$  & $7p_{1/2}$ & $8p_{1/2}$ & $6d_{3/2}$ & $7p_{3/2}$ & $8p_{3/2}$ \\
      & -0.109 & -0.066 & 0.001 & 0.000 & 0.059 & -0.005 & -0.004 \\
\end{tabular}
\end{ruledtabular}
\footnotetext[1]{$V^{N-3}$ approximation}
    \label{tab:energy}%
  \end{table}%

  \begin{table}
    \centering
    \caption{QED corrections $R_{E1}$ to the dipole matrix elements of several atoms. $R_{E1}$ is defined in equation (\ref{eq:rsp}).}
\begin{ruledtabular}
\begin{tabular}{l|rrrrr}
            Atom&\multicolumn{5}{c}{Transition} \\
\hline
&&&&&\\
Rb    & $5s$-$5p_{1/2}$ & $5s$-$6p_{1/2}$ & $6s$-$5p_{1/2}$ & $6s$-$6p_{1/2}$ & $5p_{3/2}$-$5s$ \\
      & 0.193 & -1.624 & -0.254 & 0.180 & 0.196 \\
      &       &       &       &       &  \\
Cs    & $6s$-$6p_{1/2}$ & $6s$-$7p_{1/2}$ & $7s$-$6p_{1/2}$ & $7s$-$7p_{1/2}$ & $6p_{3/2}$-$6s$ \\
      & 0.328 & -3.678 & -0.452 & 0.301 & 0.346 \\
      &       &       &       &       &  \\
Ba${}^+$ & $6s$-$6p_{1/2}$ & $6s$-$7p_{1/2}$ & $7s$-$6p_{1/2}$ & $7s$-$7p_{1/2}$ & $6p_{3/2}$-$6s$ \\
      & 0.257 & 11.75 & -0.544 & 0.240 & 0.283 \\
      &       &       &       &       &  \\
Tl\footnotemark[1]    & $6p_{1/2}$-$6s$ & $6p_{1/2}$-$7s$ & $6p_{3/2}$-$6s$ & $6p_{3/2}$-$7s$ &  \\
      & 0.619 & -0.804 & 0.717 & -0.527 &  \\
      &       &       &       &       &  \\
Fr    & $7s$-$7p_{1/2}$ & $7s$-$8p_{1/2}$ & $8s$-$7p_{1/2}$ & $8s$-$8p_{1/2}$ & $7p_{3/2}$-$7s$ \\
      & 0.647 & -5.588 & -0.731 & 0.561 & 0.768 \\
      &       &       &       &       &  \\
Ra${}^+$ & $7s$-$7p_{1/2}$ & $7s$-$8p_{1/2}$ & $8s$-$7p_{1/2}$ & $8s$-$8p_{1/2}$ & $7p_{3/2}$-$7s$ \\
      & 0.483 & 26.39 & -0.857 & 0.441 & 0.622 \\
\end{tabular}%
\end{ruledtabular}
    \label{tab:dipole}%
\footnotetext[1]{$V^{N-3}$ approximation}
  \end{table}%

Tables \ref{tab:energy} and \ref{tab:dipole} show QED corrections to the 
lowest states only. However, we include QED corrections to all states used 
in the summation (\ref{eq:pnc}).  
The results are presented in Table~\ref{tab:results}
together with the correction arising from the QED correction to the
weal matrix elements. The latter are taken from previous
works~\cite{KFQED03,MSTQED03,ueh} (see also Table~\ref{tab:weak}).

While there is typically still some cancellation between the
contributions from the energy levels and dipole matrix elements in the
transitions studied, it is not always as complete as it is with
cesium, making these results significant. 

  \begin{table}
    \centering
    \caption{QED corrections (as percentages) to PNC amplitudes for
      several atoms. Corrections due to weak matrix elements ($H_W$)
      are taken from Table~\ref{tab:weak}. Corrections due to change
      of energy denominators ($E_n$) and electric dipole transition
      amplitudes ($E1$) are the result of present work.} 
\begin{ruledtabular}
\begin{tabular}{ll|rccr}
            \multicolumn{2}{c}{Transition}& $ H_W$  & $E_n$ & $E1$ & Total \\
      \hline
&&&&&\\
      {Rb}& $5s$-$6s$ & -0.31 & -0.25 & 0.31  & -0.25(4) \\
      & $5s$-$4d_{3/2}$ & -0.31 & 0.12 & 0.001  & -0.19(5) \\
&&&&&\\
      {Cs} & $6s$-$7s$ & -0.43 & -0.42  & 0.52 & -0.33(4) \\
      {} & $6s$-$5d_{3/2}$ & -0.43 & 0.20  & -0.003 & -0.23(7) \\
&&&&&\\
      {Ba${}^+$}&  $6s$-$7s$ & -0.45 & -0.54 & 0.68  & -0.31(4) \\
      &  $6s$-$5d_{3/2}$ & -0.45 & 0.29 & -0.05  & -0.22(8) \\ 
&&&&&\\
%      {Tl}& $6p_{1/2}$-$6p_{3/2}$ & -0.50 & -0.22 & -0.18 & -0.9(2) \\
      {Tl}& $6p_{1/2}$-$6p_{3/2}$ & -0.50 & 0.07 & 0.06 & -0.37(8) \\
&&&&&\\
      {Fr}& $7s$-$8s$ & -0.60 & -0.83 & 1.02  & -0.41(8) \\
      & $7s$-$6d_{3/2}$ & -0.60 & 0.42 & -0.02  & -0.2(1) \\
&&&&&\\
      {Ra${}^+$}& $7s$-$8s$ & -0.61 & -0.97 & 1.20  & -0.38(9) \\
      & $7s$-$6d_{3/2}$ & -0.61 & 0.47 & -0.09  & -0.2(1) \\
\end{tabular}
\end{ruledtabular}
    \label{tab:results}%
  \end{table}%

Despite the fact that there are some individual corrections in the
atoms that are quite large, there is significant cancellation between
contributions from the weak matrix elements and the combined
contributions from the energies and dipoles in most of the transitions
presented. This causes the total QED contributions in most atoms to be
highly suppressed. 

The uncertainty of the total QED corrections to the PNC amplitudes
come from uncertainties of all three sources ($H_W$, $E_n$ and $E1$)
added in quadrature. The uncertainty of the first term ($H_W$) has
been estimated in~\cite{KFQED03,MSTQED03,ueh}. The estimation of
uncertainties for two other terms ($E_n$ and $E1$) comes from the
spread of values of the QED correction found in different
approximations (RPA, Brueckner, etc.).

\paragraph{Acknowledgments---}

This work was supported in part by the Australian Research Council.

%==============================================================

\end{document}